# Anomalous spin Hall magnetoresistance in Pt/Co bilayers

Masashi Kawaguchi[1], Daiki Towa[1], Yong-Chang Lau[1,2], Saburo Takahashi[3] and Masamitsu Hayashi[1,2]*

[1]*Department of Physics, The University of Tokyo, Bunkyo, Tokyo 113-0033, Japan2*

[2]*National Institute for Materials Science, Tsukuba 305-0047, Japan*

[3]*Institute for Materials Research, Tohoku University, Sendai 980-8577, Japan*

We have studied the spin Hall magnetoresistance (SMR), the magnetoresistance within the plane transverse to the current flow, of Pt/Co bilayers. We find that the SMR increases with increasing Co thickness: the effective spin Hall angle for bilayers with thick Co exceeds the reported values of Pt when a conventional drift-diffusion model is used. An extended model including spin transport within the Co layer cannot account for the large SMR. To identify its origin, contributions from other sources are studied. For most bilayers, the SMR increases with decreasing temperature and increasing magnetic field, indicating that magnon-related effects in the Co layer play little role. Without the Pt layer, we do not observe the large SMR found for the Pt/Co bilayers with thick Co. Implementing the effect of the so-called interface magnetoresistance and the textured induced anisotropic scattering cannot account for the Co thickness dependent SMR. Since the large SMR is present for W/Co but its magnitude reduces in W/CoFeB, we infer its origin is associated with a particular property of Co.

*Email: hayashi@phys.s.u-tokyo.ac.jp

The spin Hall effect[1-3] of heavy metals allows generation of spin current large enough to control the magnetization direction of a neighboring magnetic layer. The amount of spin current generated via the spin Hall effect is determined by the spin Hall angle and can vary depending on the strength of the spin orbit interaction, band filling and Berry curvature of particular orbitals, and the number of impurities that can cause spin dependent scattering. Significant effort has been made to study the spin Hall effect in heavy metal (HM)/ferromagnetic metal (FM) bilayer system as the spin current generated within the HM layer can induce magnetization switching and coherent precession, domain nucleation and domain wall motion in the FM layer.

The spin current generated by the spin Hall effect can in turn influence the charge current[4]. In particular, the resistance of bilayers consisting of a HM layer and a ferromagnetic insulator (FI) or FM layer changes depending on the magnetization direction of the FI or FM layer[5-8]. The effect, known as the spin Hall magnetoresistance (SMR), scales with the square of the spin Hall angle of the HM layer. Importantly, the way the resistance depends on the magnetization direction is distinct from that of the anisotropic magnetoresistance (AMR) that takes place in the FM layer. These features allow one to use the SMR as a convenient means to extract the magnitude of spin Hall angle of the HM layer: the spin Hall angle obtained through SMR seem to agree with that estimated by other methods[9]. Recently, studies on SMR related effects have been increasing[10]; an unidirectional component of the SMR is found in HM/FM bilayers[11], SMR has been observed in systems with antiferromagnets[12-15], the observation of the spin Nernst effect is based on the SMR theory[16-18] and the SMR analogue of the Rashba-Edelstein effect[19,20] has been found in Bi based structures[21].

The magnetoresistance (MR) of heterostructures that include Pt/Co bilayer has been under focus recently. A MR that scales with the inverse of the Co layer thickness has been reported in Pt/Co/Pt trilayers[22-24]. Such interfacial contribution to the MR was found to be anisotropic: the MR is different when the (in-plane) magnetization points along or when it is orthogonal to the current flow. Similarly proximity induced magnetic moments of the Pt layer can contribute to the MR via AMR[25] Although not specific to the Pt/Co systems, it has been reported that texture induced anisotropic scattering can influence the MR[26-28].

Here we report on a systematic study of the MR in Pt/Co bilayers. We show that an anomalously large SMR emerges in Pt/Co bilayers: the SMR increases with increasing Co thickness and the extracted effective spin Hall angle for such structures exceeds the expected value of Pt. An extended drift diffusion model that includes spin current generation and diffusion in the Co layer cannot account for the anomalous SMR. The SMR increases with decreasing temperature and increasing magnetic field for thick Co bilayers, suggesting that magnon-related effects in the Co layer has little influence on the enhancement of the SMR. The large SMR diminishes for films without the Pt layer; it is also found for W/Co but the effect is reduced for W/CoFeB. We thus infer that there is a process in Co that causes the SMR to increase with increasing Co thickness.

Samples are grown on silicon substrates coated with 100 nm thick oxide using RF magnetron sputtering. The film structure is sub.|0.5 Ta|$d_N$ Pt|$t_F$ Co|2 MgO|1 Ta (units of the thickness is nm). $D_N$ and $t_F$ represent the thickness of the HM and FM layers, respectively. A 0.5 nm thick Ta layer is used as a seed layer for the growth of the Pt layer. The MgO|Ta is used as a capping layer. Hall bars are created via shadow masking the deposition. The width of the Hall bars is ~0.4 mm and the distance between the voltage probes to measure the resistance is ~1.2-1.5 mm. Transport measurements are carried out in He-filled cryostat (Quantum Design, Physical Properties Measurement System).

The inset of Fig. 1(a) shows a schematic of the film structure together with the definition of the coordinate system. Current is passed along the x-direction. The Pt and Co thickness dependence of the resistivity is plotted in Figs. 1(a) and 1(b), respectively. The resistivity is obtained by measuring the layer thickness dependence of the sheet conductance ($G_{XX}$). The fluctuation in the resistivity for thin Pt and/or Co layer films arises since the resistivity is estimated from the thickness derivative of the sheet conductance. (See the Supplementary Material (Fig. S1) for the details of the extraction of the resistivity.) The resistivity of the Pt and Co layers shows a significant thickness dependence most likely due to increased interface scattering when the thickness is reduced[6,23,29].

The MR of the Pt/Co bilayers are shown in Fig. 2. We measure the resistance ($R_{XX}$) of the Hall bars while rotating the magnetic field in the y-z plane. We define the angle between the magnetic field (H) and the z-axis within the y-z plane as $\varphi$ (see Fig. 1(a)). The $\varphi$ dependence of $R_{XX}$ is fitted with a sinusoidal function $R_{XX} = R_0 \left[1 - \frac{\Delta R_{y-z}}{2} \cos(2\varphi + \varphi_0)\right]$. From hereafter, we refer to $\Delta R_{y-z}$ as the SMR even though it may contain contributions from other MR effects. $\Delta R_{y-z}$ is plotted as a function of the Pt thickness ($d_{Pt}$) and Co thickness ($t_{Co}$) in Figs. 2(a-e) and 2(f-i), respectively. The phase offset $\varphi_0$ is nearly zero for all films. $R_0$ is used to estimate the sheet conductance $G_{XX}$.

For a given Co thickness, the characteristics of the Pt thickness dependence of $\Delta R_{y-z}$ is similar to what has been reported previously in similar systems[5-8]: $|\Delta R_{y-z}|$ takes a maximum and gradually decreases with increasing $d_{Pt}$. The trend is in agreement with the drift diffusion model established to describe the spin Hall magnetoresistance[4,8] and can be considered as a fingerprint of SMR. In contrast, the Co thickness dependence of $\Delta R_{y-z}$ is different from what the drift diffusion model predicts. $\Delta R_{y-z}$ increases with increasing $t_{Co}$ and the Co thickness at which $|\Delta R_{y-z}|$ saturates depends on $d_{Pt}$. According to the model, however, $|\Delta R_{y-z}|$ should roughly scale with $1/t_{Co}$ due to current shunting into the Co layer.

To illustrate the unusual characteristics of the $\Delta R_{y-z}$ found in the Pt/Co bilayers, we fit the Pt thickness dependence of the normalized $\Delta R_{y-z}$ with the drift-diffusion model. $\Delta R_{y-z}$ is normalized by a factor ($x_N$) that takes into account current shunting into the Co layer: $x_N = (d_N \rho_F)/(d_N \rho_F + t_F \rho_N)$, where $\rho_N$ ($\rho_F$) and $d_N$ ($t_F$) represent the resistivity and thickness of the HM (FM) layers, respectively. (Note that $N$=Pt and $F$=Co in the experiments.) $\Delta R_{y-z}/x_N$ is plotted as a function of $d_N$ in Fig. 3(a). In the drift-diffusion model, here we assume transparent HM/FM interface and zero longitudinal spin absorption into the FM layer[8], which returns the lower limit of the effective spin Hall angle $\theta_{SH}$. The equation used for the fitting is:

$$\frac{\Delta R_{y-z}}{x_N} = -\theta_{SH}^2 \frac{\lambda_N}{d_N} \tanh\left(\frac{d_N}{2\lambda_N}\right)\left[1 - \frac{1}{\cosh(d_N/\lambda_N)}\right] \quad (1)$$

where $\lambda_N$ is the spin diffusion length of the HM layer. $\lambda_N$ and $\theta_{SH}$ are the fitting parameters and the resulting fitted curves are shown in Fig. 3(a). The agreement between the experimental results and Eq. (1) are quite good.

In Fig. 3(b) and 3(c), we show the Co thickness dependence of $\lambda_N$ and $\theta_{SH}$ obtained from the fitting. We find that $\lambda_N$ shows little dependence on $t_{Co}$ whereas $\theta_{SH}$ increases with increasing $t_{Co}$. The estimated $\theta_{SH}$ does not seem to saturate with Co thickness of ~7 nm and the maximum value obtained at $t_{Co}$~7 nm exceeds the effective spin Hall angle of Pt reported thus far[30-32]. We have also studied the FM layer thickness dependence of HM/FM bilayers with HM=W and found again an extraordinary large SMR ($\Delta R_{y-z}/x_N$) only when the FM layer is Co (see Supplementary material Fig. S2): $\Delta R_{y-z}/x_N$ matches the value of the drift-diffusion model for FM=CoFeB, as reported previously[8,9,33]. We thus consider the large $\Delta R_{y-z}/x_N$ found in Pt/Co bilayers is likely due to bulk spin transport in the Co layer.

Since the SMR increases with increasing Co thickness, one may consider spin current generation within the Co layer via excitations of magnons. Magnon is the main carrier of spin current in ferromagnetic insulators (FI) and causes the spin Seebeck effect in HM/FI structures[34]. Any magnon related effect, however, will be suppressed by application of large magnetic field or by lowering the measurement temperature[35]. We have thus studied the field and temperature dependence of the SMR of Pt/Co bilayers for selected films.

Figure 4(a) shows $\Delta R_{y-z}$ plotted as a function of the magnitude of the magnetic field $|H|$. Here we have measured the bilayer resistance against the magnetic field directed along the $y$- ($R_{XX}^Y$) and $z$- ($R_{XX}^Z$) axes and took their difference, i.e., $\Delta R_{y-z}=(R_{XX}^Y-R_{XX}^Z)/R_{XX}^Z$. The definition of $\Delta R_{y-z}$ is similar to that described above using the sinusoidal function if one assumes $R_{XX}^Z \sim R_0$. In all cases, we find $|\Delta R_{y-z}|$ slightly increases with increasing $|H|$. (For $|H|$ smaller than ~15 kOe, the hard axis magnetic field is not sufficient to force the magnetization to point along the field direction, thus $\Delta R_{y-z}$ decreases with decreasing $|H|$.) We consider the slight increase in $|\Delta R_{y-z}|$ with increasing $|H|$ above ~15 kOe may originate from contribution of the so-called Hanle magnetoresistance[36,37], which causes an increase in $R_{XX}^Z$ with increasing $|H|$ and thus results in enhancement of $|\Delta R_{y-z}|$. Although the Hanle magnetoresistance may influence $|\Delta R_{y-z}|$, we do not find significant reduction of $|\Delta R_{y-z}|$ at large $|H|$ similar to that reported for the magnetic field dependence of the spin Seebeck coefficient in Pt/YIG[35].

The temperature dependence of $\Delta R_{y-z}$ is plotted in Fig. 4(b). Except for Pt/Co bilayer with $d_N \sim 8$ nm and $t_F \sim 1$ nm, $|\Delta R_{y-z}|$ increases with decreasing temperature. The change is more pronounced for films with larger $t_F$ and/or $d_N$. The temperature dependence of $\Delta R_{y-z}$ is similar to what has been found in metallic bilayers: $|\Delta R_{y-z}|$ increases with decreasing temperature which has been attributed to the change in the degree of spin absorption[8]. Since we do not find any reduction in $|\Delta R_{y-z}|$ at low temperature, we consider any magnon-related effect is almost negligible for the anomalously large SMR for thicker Co bilayers.

As $|\Delta R_{y-z}|$ becomes larger when the Co thickness is increased, we infer the enhancement is due to an effect that takes place in the bulk of the Co layer. Since the spin diffusion length of Co is reported to be much larger than the film thickness used here[38], it may be feasible to consider spin current generation within the Co layer via, e.g. anomalous Hall and/or spin Hall effects. One can include contributions from the anomalous Hall and spin Hall effects of the FM layer (as well as longitudinal and transverse spin-current absorption) into the drift-diffusion model and estimate the resulting SMR ($R_{\mathrm{SMR}}$).

$$R_{\mathrm{SMR}} = -(1-x_N)x_N\theta_{AH}^2 - (1-x_N)(1-P^2)\theta_F^2 \frac{2\lambda_F}{t_F}\tanh\left(\frac{t_F}{2\lambda_F}\right)$$

$$- x_N\theta_{SH}^2 \frac{\lambda_N}{d_N}\tanh^2\left(\frac{d_N}{2\lambda_N}\right)\left\{\mathrm{Re}\left[\frac{g_S}{1+g_S\coth(d_N/\lambda_N)}\right] - \frac{(1+\zeta)^2 g_F}{1+g_F\coth(d_N/\lambda_N)}\right\} \quad (2)$$

where $g_S = \rho_N\lambda_N G_{\mathrm{MIX}}$, $g_F = (1-P^2)\frac{\rho_N\lambda_N}{\rho_F\lambda_F}\tanh\left(\frac{t_F}{\lambda_F}\right)$, $\zeta = \frac{\theta_F\lambda_F\tanh(t_F/2\lambda_F)}{\theta_{SH}\lambda_N\tanh(d_N/2\lambda_N)}$. $G_{\mathrm{MIX}}$ is the spin mixing conductance of the HM/FM interface, $\theta_F$, $\theta_{AH}$, $\lambda_F$ and $P$ are the spin Hall angle, anomalous Hall angle, spin diffusion length and spin polarization of the FM layer. The first and second terms of the right-hand side of Eq. (2) represent contributions to $\Delta R_{y-z}$ from the anomalous Hall and spin Hall effects of the FM layer, respectively. The second term in the curly bracket represents the longitudinal spin absorption contribution to the SMR[8]. This term is modified ($\zeta \neq 0$) when the FM spin Hall effect is considered. With $\theta_F=\theta_{AH}=0$, $\text{Re}[G_{MIX}] \gg \text{Im}[G_{MIX}]$ and $\text{Re}[G_{MIX}] \gg 1/(2\rho_N\lambda_N)$, we recover Eq. (1). The transverse spin diffusion length[39,40] of the FM layer is assumed to be zero here.

Representative values of the calculated $R_{\text{SMR}}$ are shown in the supplementary material (Fig. S4). In the calculations, the spin Hall angle of Pt is set to one of the largest values reported thus far ($\theta_{SH} \sim 0.2$)[30], the anomalous Hall angle is obtained from the measurements (the largest value of the Pt/Co bilayers is ~0.03, see Fig. S3) and the spin Hall angle of Co is estimated using the relationship $\theta_{AH} \sim P\theta_F$: here we use $P \sim 0.4$. Although the magnitude of $R_{\text{SMR}}$ can be adjusted to match the experimental results of $\Delta R_{y-z}$, it is difficult to reproduce the characteristics found in the experiments shown in Fig. 2, the increasing $\Delta R_{y-z}$ with increasing Co thickness.

We have also studied the MR within the *x-z* plane which provides information on the so-called anisotropic interface magnetoresistance[22-24] (for clarity of notation, we denote such MR as IMR instead of AIMR used in Ref. [22]). $R_{XX}$ is measured while the magnetic field is rotated in the *x-z* plane. The angle between the magnetic field and the *z*-axis within the *x-z* plane is defined as $\gamma$. We fit the $\gamma$ dependence of $R_{XX}$ using a sinusoidal function $R_{XX} = R_0 \left[1 - \frac{\Delta R_{x-z}}{2}\cos(2\gamma + \gamma_0)\right]$. $\Delta R_{x-z}$ is plotted as a function of Co thickness ($t_{\text{Co}}$) in Fig. 5(a). The large $t_{\text{Co}}$ limit of $\Delta R_{x-z}$ normalized by the current flowing in the Co layer, i.e. $\frac{\Delta R_{x-z}}{1-x_N}$, gives the anisotropic magnetoresistance (AMR) of bulk Co ($R_{\text{AMR}}$) whereas changes at smaller $t_{\text{Co}}$ reflect contribution from the interface(s). In Fig. 5(b), we plot $\frac{\Delta R_{x-z}}{1-x_N}$ vs. $1/t_{\text{Co}}$. The *y*-axis intercept of the linear fitting in the appropriate $t_{Co}$ range, as shown in Fig. 5(b), gives $R_{AMR}$. We find that $R_{AMR}$ for films with different Pt thickness are nearly the same ($R_{AMR} \sim 1.6\%$), which is close to what has been reported previously[41]. IMR is obtained by the following relation, $R_{IMR} \equiv \frac{\Delta R_{x-z}}{1-x_N} - R_{AMR}$. The Pt thickness dependence of $R_{IMR}$ is plotted in Fig. 5(c). As reported previously[21,23,24], $R_{IMR}$ increases with decreasing $t_{Co}$.

Although the microscopic mechanism is not clear, one may assume[23,24] that $R_{IMR}$ is isotropic within the film plane, i.e. $R_{IMR}$ contributes to the *y-z* plane MR in the same way it does in the *x-z* plane. Assuming a parallel circuit that includes contribution from the SMR within the HM layer ($R_{SMR}$) and the IMR that takes place in the FM layer ($R_{IMR}$), we obtain an expression for $\Delta R_{y-z}$, which can be rewritten as: $R_{SMR} = \Delta R_{y-z} \frac{1}{x_N} - (R_{IMR} + \Delta R_{y-z}^{Co}) \frac{t_F}{\rho_F} \frac{\rho_N}{d_N}$ (see Supplementary material for the derivation). Here we have also subtracted $\Delta R_{y-z}$ for films with $d_{Pt}=0$, which is denoted as $\Delta R_{y-z}^{Co}$. $\Delta R_{y-z}^{Co}$ is not related to SMR and is likely associated with the textured induced anisotropic scattering (also referred to as the geometrical size effect[27]). In Fig. 5(d), $R_{SMR}$ is plotted as a function of $d_{Pt}$. As evident, $|R_{SMR}|$ increases with increasing Co thickness, which cannot be described by the extended drift diffusion model (including spin transport within the FM layer), i.e. Eq. (2). Moreover, the sign of $R_{SMR}$ reverses when $d_{Pt} \sim 2$-3 nm and its magnitude increases with decreasing $d_{Pt}$: see the inset of Fig. 5(d). Such trend of $R_{SMR}$, inconsistent with the $d_{Pt}$ dependence of SMR, is largely caused by the subtraction of the $R_{IMR}$ term. These results show that the large $\Delta R_{y-z}$ which increases with the Co thickness cannot be described by combinations of SMR, an isotropic IMR and texture induced anisotropic scattering.

Since $\Delta R_{y-z}$ increases with $t_{Co}$, interface effects (e.g. spin memory loss[42], proximity magnetization induced AMR[25], spin orbit interface AMR[43,44], Edelstein magnetoresistance[20,21], enhanced spin Hall effect at the interface[45]) may not play a dominant role in setting the SMR found here. It is also difficult to foresee thermal effects (e.g. anomalous Nernst effect[46]) give rise to the large SMR since we find nearly symmetric resistance levels for positive and negative magnetic field in all field directions[17]. It has been suggested that the magnetoresistance due to texture induced anisotropic scattering in the FM layer can be enhanced by attaching a NM layer with large spin orbit interaction (the length scale of such MR is associated with the electron mean free path of the FM layer)[28]. With respect to the texture of the Co layer, we find a large $\Delta R_{y-z}$ both in Pt/Co and W/Co bilayers despite the different texture of the HM underlayer. Furthermore, $\Delta R_{y-z}$ of Ta/Co bilayers (i.e. the case when $d_{Pt}=0$) shows a much reduced $\Delta R_{y-z}$ compared to that of W/Co bilayers. Since Ta and W are both amorphous-like[9], the texture of the Co layer is expected to be similar for the two bilayers. We thus infer that the texture of the Co film has little to do with the large $\Delta R_{y-z}$. It is possible that the spin orbit coupling at the interface may be different for the three HM layers (Pt, W and Ta), as manifested by the difference in the Dzyaloshinskii-Moriya interaction[47,48], and may influence the MR by means of anisotropic scattering. As the microscopic origin of the such effect remains to be identified, further investigation is required to clarify its presence.

In summary, we have studied the Pt and Co thickness dependence of spin Hall magnetoresistance), the magnetoresistance within the plane transverse to the current flow, in Pt/Co bilayers. We find SMR that increases with increasing Co thickness, which cannot be fully accounted for even if the anomalous Hall and spin Hall effects of the Co layer are considered. Other effects that may contribute to the MR are evaluated to assess their influence on the SMR. We find that the so-called anisotropic interface magnetoresistance and the magnetoresistance due to texture induced anisotropic scattering cannot describe the large SMR observed here. Further investigation is required to clarify the origin of the SMR in HM/Co bilayers.

**Supplementary Material**

The following topics are included in the supplementary material: description on how the thickness dependent resistivity is obtained, additional experimental results on the anomalous Hall resistance of Pt/Co bilayers and SMR of W/Co and W/CoFeB bilayers, model calculations of the SMR using the drift-diffusion model, and description of how IMR is extracted.

**Acknowledgements**

We thank S. Mitani for fruitful discussion. This work was partly supported by JSPS Grant-in-Aids for Specially Promoted Research (15H05702), Casio Foundation, and the Center of Spintronics Research Network of Japan. Y.-C.L. is a JSPS international research fellow.

**Figure captions**

**Figure 1**. (a) Pt layer thickness dependence of the resistivity of Pt ($\rho_{Pt}$) and (b) Co layer thickness dependence of the resistivity of Co ($\rho_{Co}$) in Pt/Co bilayer. Lines are guide to the eyes. The inset to (a) shows a schematic view of the system and definition of the coordinate axis.

**Figure 2**. (a) Magnetoresistance in the *y-z* plane ($\Delta R_{y-z}$) plotted as a function of the Pt layer thickness ($d_{Pt}$) for fixed Co layer thicknesses. (f-j) $\Delta R_{y-z}$ vs. Co layer thickness ($t_{Co}$) for fixed Pt layer thicknesses. The thickness of the fixed thickness layer is noted in each panel.

**Figure 3**. (a) $\Delta R_{y-z}/x_N$ plotted as a function of Pt thickness. The thickness of the Co layer for each symbol is displayed in the legend. Solid lines show fitting results using Eq. (1). (b,c) Spin diffusion length $\lambda_N$ (b) and the spin Hall angle $\theta_{SH}$ (c) of Pt obtained from the fitting shown in (a).

**Figure 4.** (a,b) Applied magnetic field (a) and the measurement temperature (b) dependence of $\Delta R_{y-z}$ for representative Pt/Co bilayers. The film structure is described in the legend.

**Figure 5**. (a) Magnetoresistance in the *x-z* plane ($\Delta R_{x-z}$) plotted as a function of the Co layer thickness ($t_{Co}$) for fixed Pt layer thickness ($d_{Pt}$). (b) $\Delta R_{x-z}/(1-x_N)$ vs. $1/t_{Co}$. The solid lines show linear fit to the data when $1/t_{Co}$ is small. The *y*-axis intercept of the linear line gives the AMR of bulk Co ($R_{AMR}$). (c) Interface magnetoresistance $R_{IMR} \equiv \frac{\Delta R_{x-z}}{1-x_N} - R_{AMR}$ plotted as a function of $d_{\text{Pt}}$. (d) $R_{\text{SMR}} = \Delta R_{y-z}\frac{1}{x_N} - \left(R_{\text{IMR}} + \Delta R_{y-z}^{\text{Co}}\right)\frac{t_F}{\rho_F}\frac{\rho_N}{d_N}$ vs. $d_{\text{Pt}}$. The inset shows the same plot but with the vertical axis range expanded.

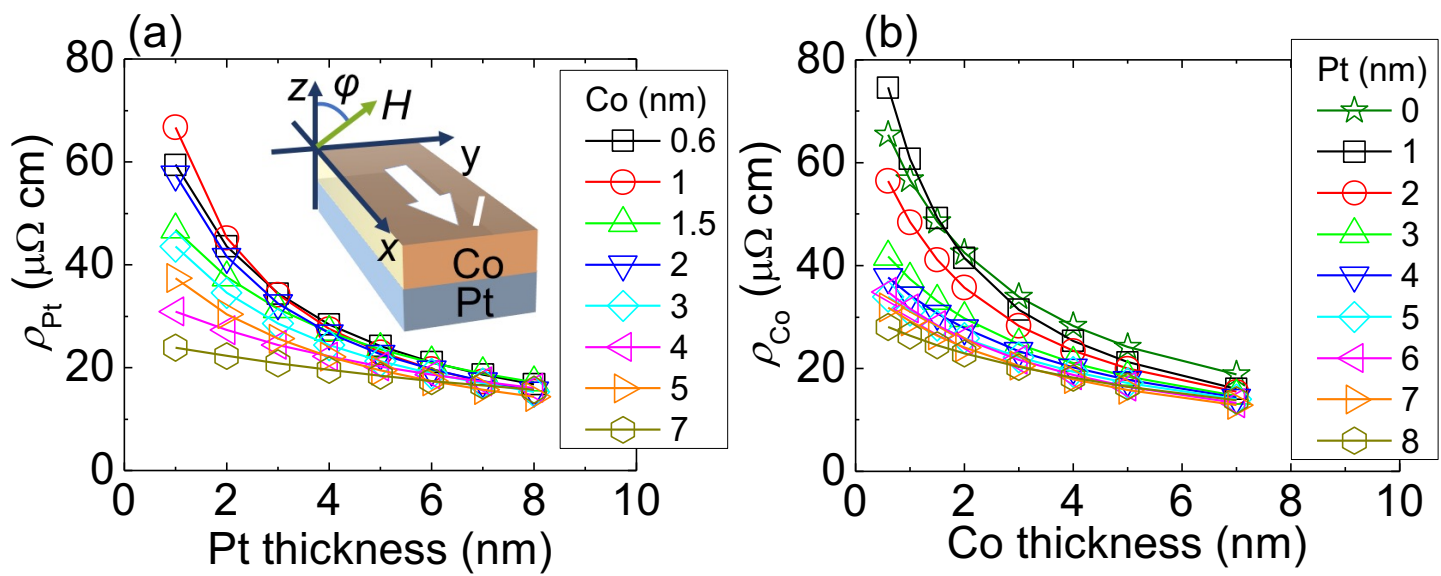

Fig. 1

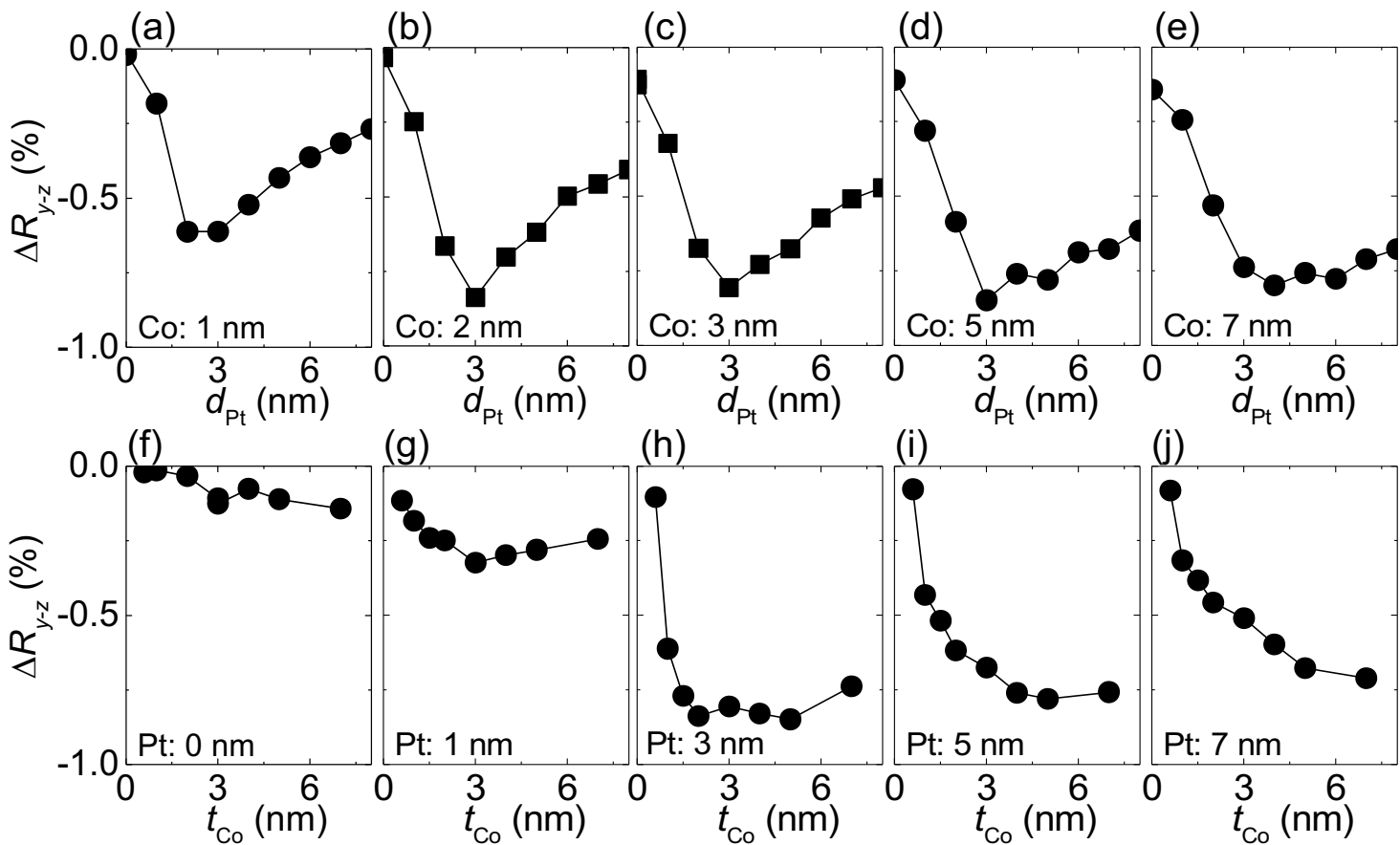

Fig. 2

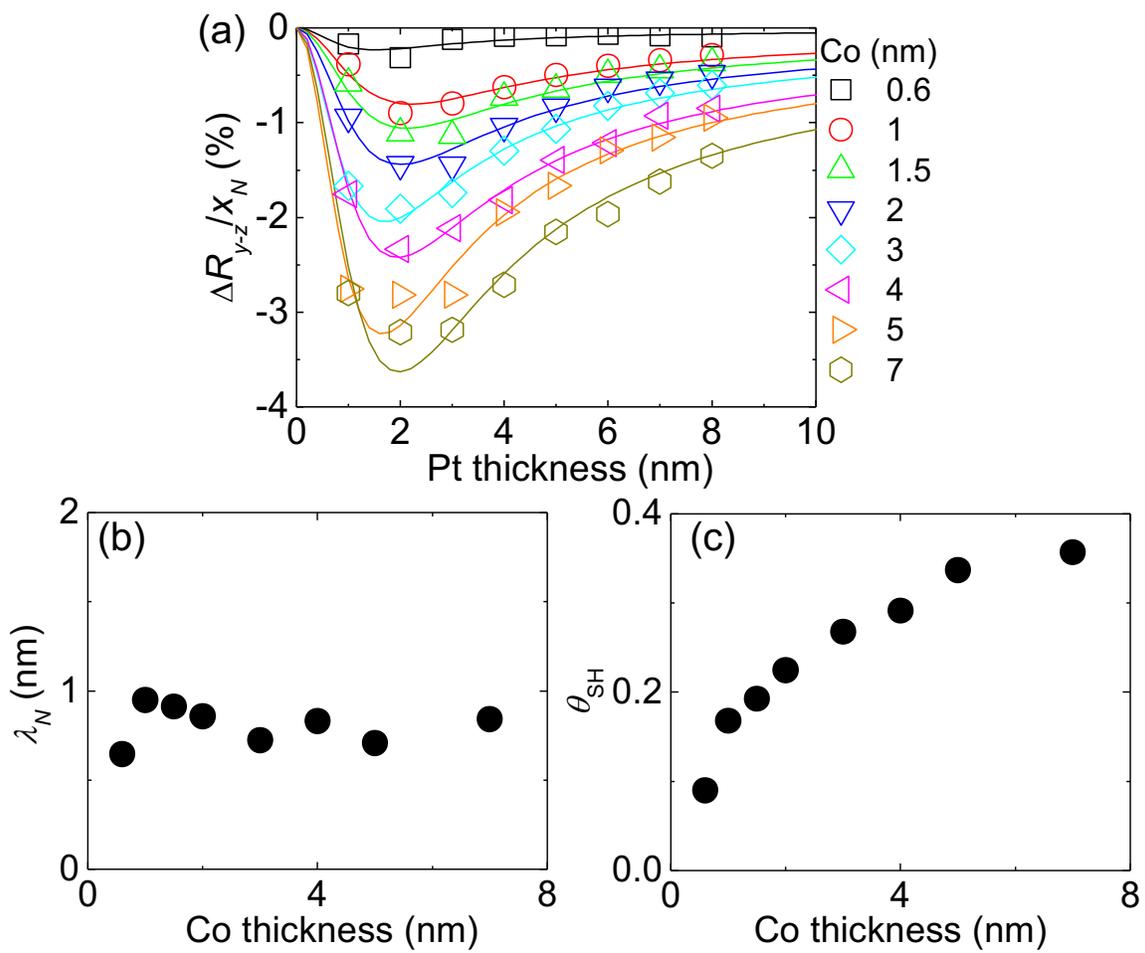

Fig. 3

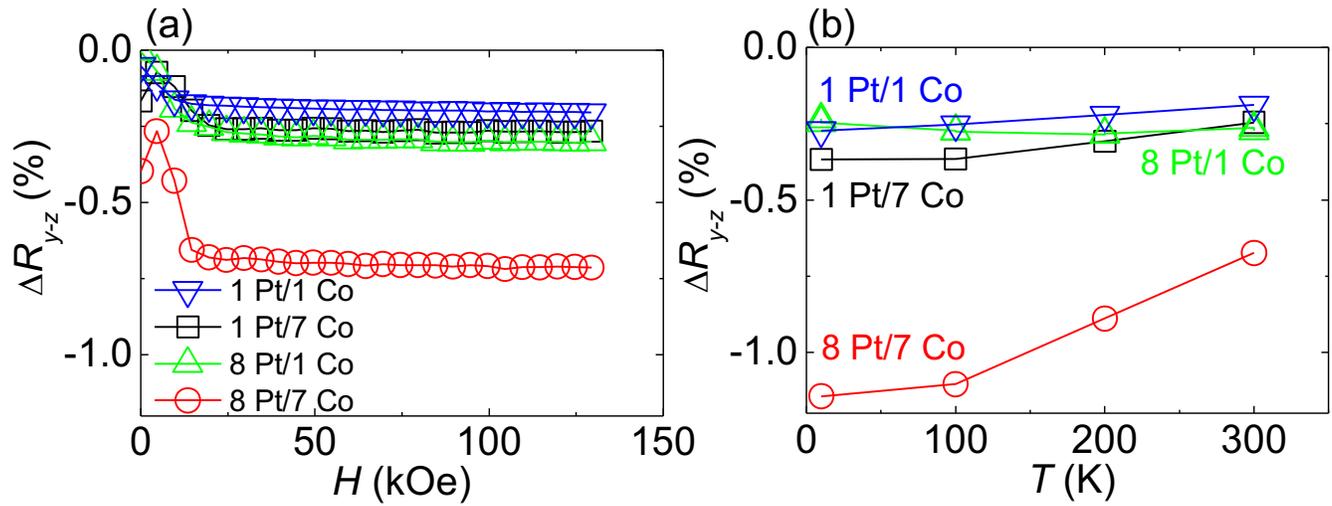

Fig. 4

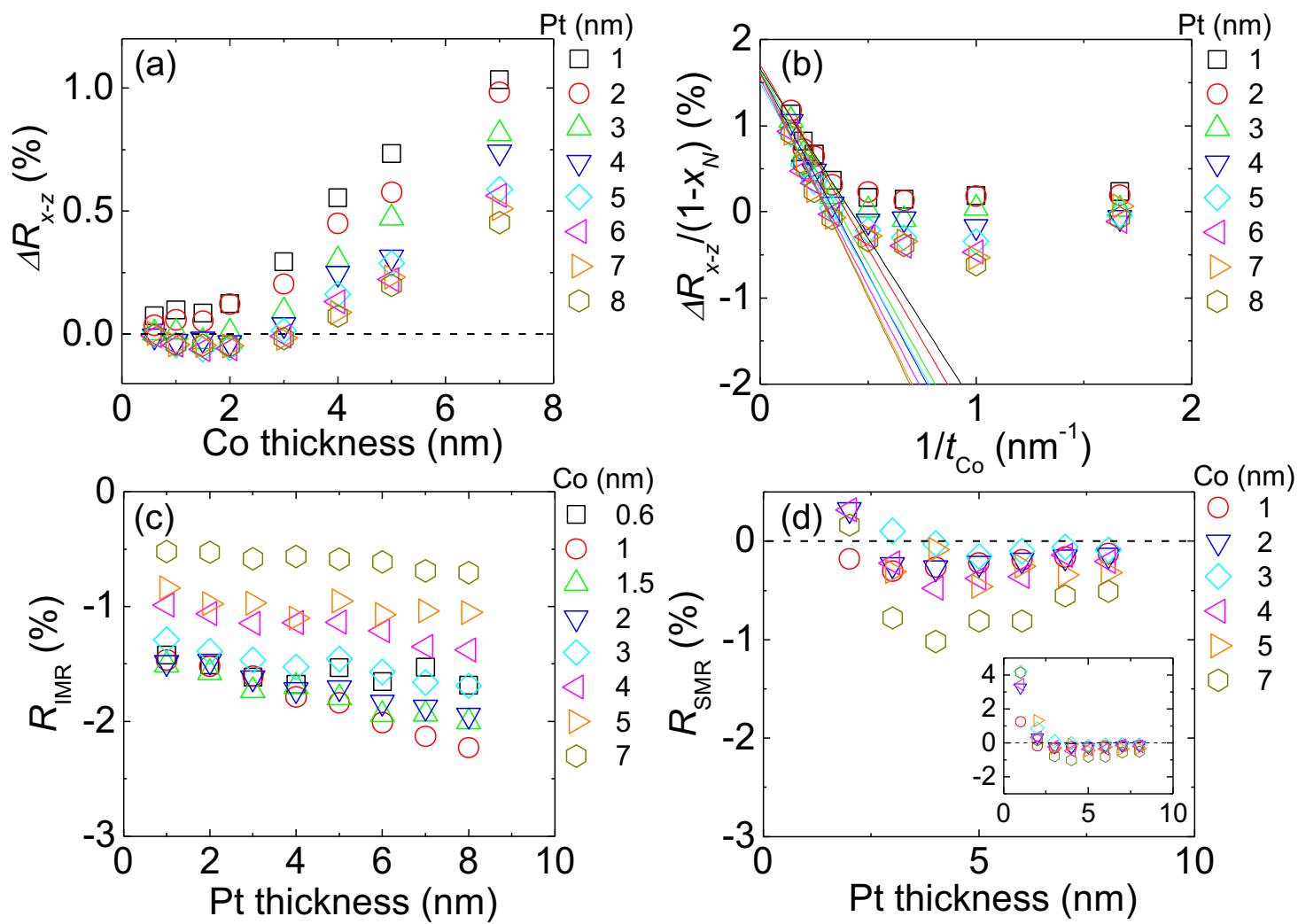

Fig. 5

Supplementary material for

# Anomalous spin Hall magnetoresistance in Pt/Co bilayers

Masashi Kawaguchi[1], Daiki Towa[1], Yong-Chang Lau[1,2], Saburo Takahashi[2] and Masamitsu Hayashi[1,2]*

[1]*Department of Physics, The University of Tokyo, Bunkyo, Tokyo 113-0033, Japan2*

[2]*National Institute for Materials Science, Tsukuba 305-0047, Japan*

[3]*Institute for Materials Research, Tohoku University, Sendai 980-8577, Japan*

## S1. Extraction of the thickness dependent resistivity

The sheet conductance $G_{XX}$ is evaluated for each sample in order to calculate distribution of the current within the bilayer. We use four-point probe technique to measure the resistance $R_0$ (see the main text for the definition of $R_0$) of a Hall bar (width: $w$, distance between the voltage probes: $L$). The sheet conductance is defined as $G_{XX}=L/(wR_0)$. Figure S1 shows the sheet conductance of the Pt/Co bilayers as a function of Pt ($d_N$) and Co ($t_F$) thicknesses. Clearly $G_{XX}$ depends on $t_F$ and $d_N$. We use second degree polynomial fitting to extract the thickness dependent intrinsic resistivity. Note that the Fuchs-Sondheimer model gives such dependence when the layer thickness is smaller than its mean free path. The solid lines in Fig. S1 shows the fitted curves $f(d_N,t_F)$ using a parabolic function. To determine the conductivity of each layer as a function of $d_N$ or $t_F$, we take the thickness derivative of $f(d_N,t_F)$ and define the conductivities as follows:

$$\sigma_N(t_F, d_N) \equiv \frac{\partial f(t_F, d_N)}{\partial d_N} \qquad (S1)$$

$$\sigma_F(t_F, d_N) \equiv \frac{\partial f(t_F, d_N)}{\partial t_F} \qquad (S2)$$

Here, $\sigma_N$ and $\sigma_F$ are the conductivity of Pt and Co, respectively. The resistivity is obtained using the following relations: $\rho_{Pt} \sim 1/\sigma_N$ and $\rho_{Co} \sim 1/\sigma_F$ (here we have neglected the transverse component of the conductivity). $\rho_{Pt}$ and $\rho_{Co}$ are displayed in Figs. 1(b) and 1(c), respectively.

## S2. W/Co and W/CoFeB bilayers

To check whether the combination of Pt and Co is essential to observing the unusually large SMR, we have studied the FM layer thickness dependence of HM/FM bilayers with HM=W and FM=Co and CoFeB. The films are made by RF magnetron sputtering. The film structure is sub.|3 W| $t_F$ Co|2 MgO|1 Ta and sub.|3 W| $t_F$ CoFeB|2 MgO|1 Ta (units of the thickness is nm). Films are annealed at 300 °C for 1 hour ex-situ in vacuum. Results from these films are presented in Fig. S2.

## S3. Anomalous Hall resistance

The Pt layer thickness dependence of the anomalous Hall resistance $R_{AHE}$ is shown in Fig. S3. $R_{AHE}$ is the difference in the Hall resistance $R_{XY}$ when the magnetization of the Co layer points along $+z$ and $-z$. $R_{AHE}$ is divided by the sheet resistance $R_{XX}$ and $(1-x_N)$: the anomalous Hall angle $\theta_{AH}$ of the FM layer corresponds to $-R_{AHE}/R_{XX}/(1-x_N)$ if one neglects contribution of the diffusive spin current from the Pt layer on the Hall resistance[1].

## S4. Model calculations

The Pt and Co layer thickness dependence of $R_{SMR}$ calculated using Eq. (2) of the main text is shown in Fig. S4, black solid lines. The following parameters are used in the calculations. Co layer: $\theta_F$=0.075, $\theta_{AH}$=0.03, $P$=0.4$^2$, $\lambda_F$=40 nm (ref. [2]). Pt layer: $\theta_{SH}$=0.2 (ref. [3]), $\lambda_N$=1 nm. The resistivity of each layer is taken from Fig. 1. Interface: Re[$G_{MIX}$]=10$^{14}$ $\Omega^{-1}$cm$^{-2}$, Im[$G_{MIX}$]=10$^{10}$ $\Omega^{-1}$cm$^{-2}$. Note that such $G_{MIX}$ represents a transparent Pt/Co interface for spin transmission. To illustrate the effect of spin transport in the Co layer on the SMR, the blue solid lines in Fig. S4 show $R_{SMR}$ calculated with the same parameters as above but with $\theta_F$=0 and $\theta_{AH}$=0.

## S5. Influence of IMR on SMR

In order to evaluate how IMR influences the SMR, we subtract IMR from $\Delta R_{y-z}$ and extract SMR with the assumption that IMR in the $y$-$z$ plane has the same magnitude with that of $x$-$z$ plane. To model the system, we assume that the resistance change due to SMR occurs in the HM layer (e.g. Pt) whereas the change caused by the IMR takes place within the FM layer (e.g. Co). One can therefore consider a parallel circuit that describes the magnetoresistance due to SMR and IMR as the following. We define the change in the current when the magnetic field is applied along the $y$-axis and $z$-axis as $\Delta I$. As noted above, $\Delta I$ can be divided into two components, changes that occur in the FM layer ($\Delta I_{FM}$) and the HM layer ($\Delta I_{HM}$).

From the parallel circuit model, we obtain

$$\Delta I = \Delta I_{HM} + \Delta I_{FM} \qquad (S3)$$

$\Delta I$ can be expressed using the average film resistivity $\rho$ and its change $\Delta \rho$ due to the field application along $y$ and $z$, film thickness $t=d_N+t_F$ and the voltage $V$ applied across the films as:

$$\Delta I = Vt\left(\frac{1}{\rho+\Delta\rho} - \frac{1}{\rho}\right) \qquad (S4)$$

Similarly, using the parameters defined in the main text, we obtain

$$\Delta I_{HM} = Vd_N\left(\frac{1}{\rho_N+\Delta\rho_N} - \frac{1}{\rho_N}\right) \qquad (S5)$$

$$\Delta I_{FM} = Vt_F\left(\frac{1}{\rho_F+\Delta\rho_F} - \frac{1}{\rho_F}\right) \qquad (S6)$$

Note that here we assume contribution from the transverse resistivity on the longitudinal current is negligible. We define the resistivity change ratio in the HM and the FM layer as $R_{SMR} = \Delta\rho_N/\rho_N$ and $R_{IMR} + \Delta R_{y-z}^{Co} = \Delta\rho_F/\rho_F$, respectively, and the average ratio within the film as $\Delta R_{y-z} = \Delta\rho/\rho$. Equations (S3)-(S6) can be put together to read:

$$\frac{\Delta R_{y-z}}{(1+\Delta R_{y-z})}\left(\frac{t_F}{\rho_F} + \frac{d_N}{\rho_N}\right) = \frac{R_{SMR}}{(1+R_{SMR})}\frac{d_N}{\rho_N} + \frac{R_{IMR}+\Delta R_{y-z}^{Co}}{(1+R_{IMR}+\Delta R_{y-z}^{Co})}\frac{t_F}{\rho_F} \qquad (S7)$$

Assuming $\Delta R_{y-z}, R_{SMR}, R_{IMR} \ll 1$, we obtain

$$\Delta R_{y-z}\left(\frac{t_F}{\rho_F} + \frac{d_N}{\rho_N}\right) = \left(R_{IMR} + \Delta R_{y-z}^{Co}\right)\frac{t_F}{\rho_F} + R_{SMR}\frac{d_N}{\rho_N} \qquad (S8)$$

Rearranging Eq. (S8) will give the relation described in the main text.

**Figure captions**

**Figure S1**. (a) Pt layer thickness dependence of the sheet conductance $G_{XX}$ and (b) Co layer thickness dependence of $G_{XX}$ in Pt/Co bilayers. Lines are fit to the data using a parabolic function.

**Figure S2**. (a,b) Ferromagnetic metal (FM) layer thickness dependence of the sheet conductance ($G_{XX}$) of HM/FM bilayers. The heavy metal (HM) layer is W. FM=Co (a) and FM=CoFeB (b). The solid lines show fit to the data in selected FM thickness ranges. Resistivity ($\rho_F$) extracted from the fitting are displayed in each panel. (c,d) $\Delta R_{y-z}$ (solid squares) and $\Delta R_{y-z}/x_N$ (open circles) vs. FM layer thickness for W/FM bilayers. FM=Co (c) and FM=CoFeB (d). Lines are guide to the eyes.

**Figure S3**. (a-e) Pt ($d_{Pt}$) layer thickness dependence of the anomalous Hall resistance $R_{AHE}$ divided by the sheet resistance $R_{XX}$ and (1-$x_N$). The thickness of the Co layer is indicated in each panel.

**Figure S4**. (a-e) Calculated spin Hall magnetoresistance ($R_{SMR}$) plotted as a function of the Pt layer thickness ($d_{Pt}$) for fixed Co layer thicknesses. (f-j) Calculated $R_{SMR}$ vs. Co layer thickness ($t_{Co}$) for fixed Pt layer thicknesses. The thickness of the fixed thickness layer is noted in each panel. Parameters used in the calculations are described in the text. The black solid lines show $R_{SMR}$ calculated using Eq. (2) with the parameters described in the text. The blue solid lines are calculated similarly except for $\theta_F$ and $\theta_{AH}$, which are both set to 0.

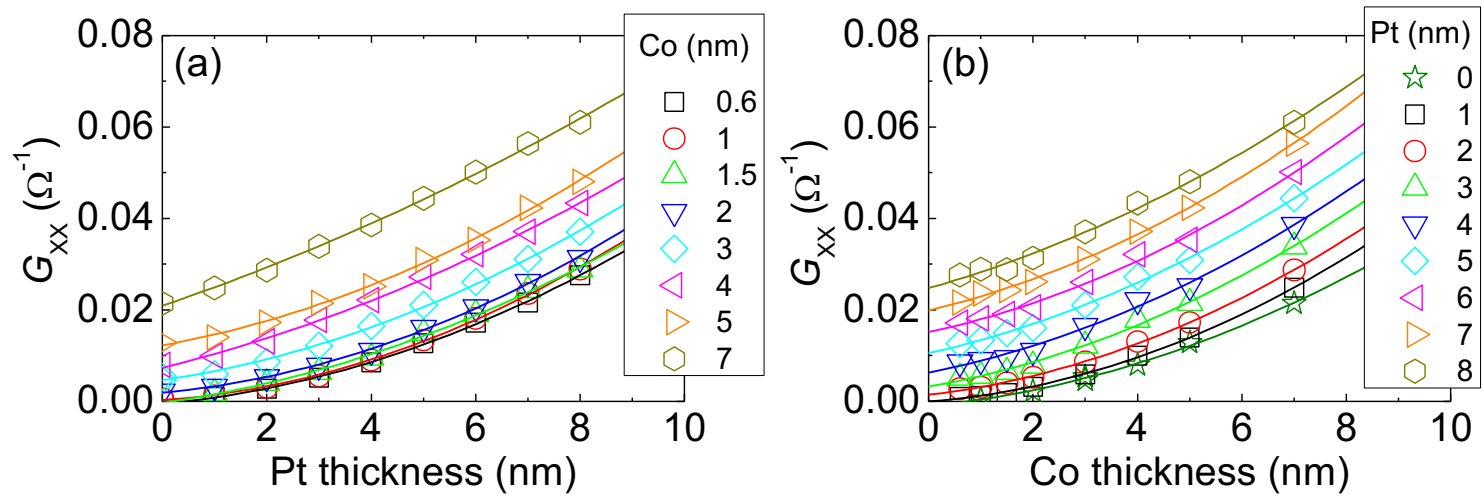

Fig. S1

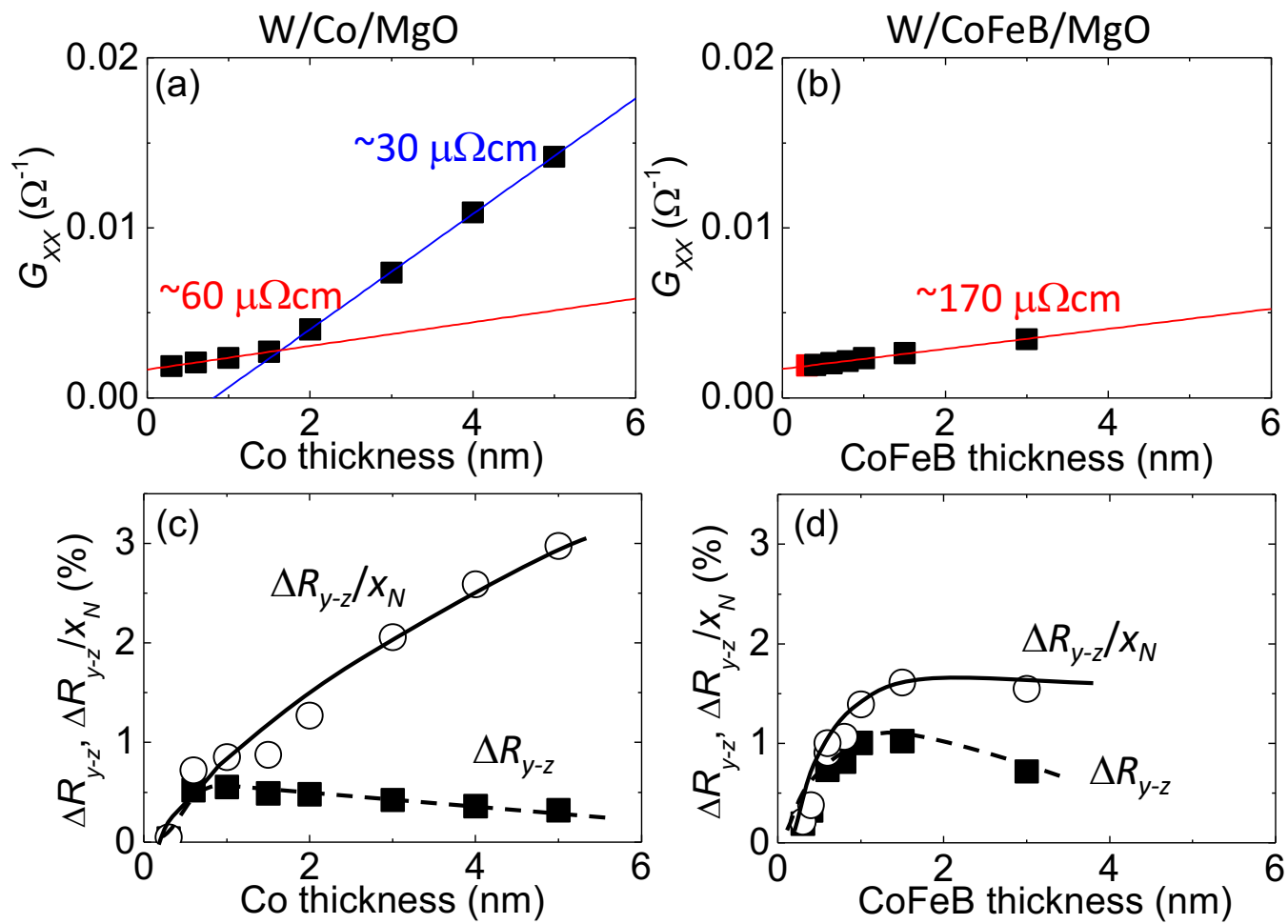

Fig. S2

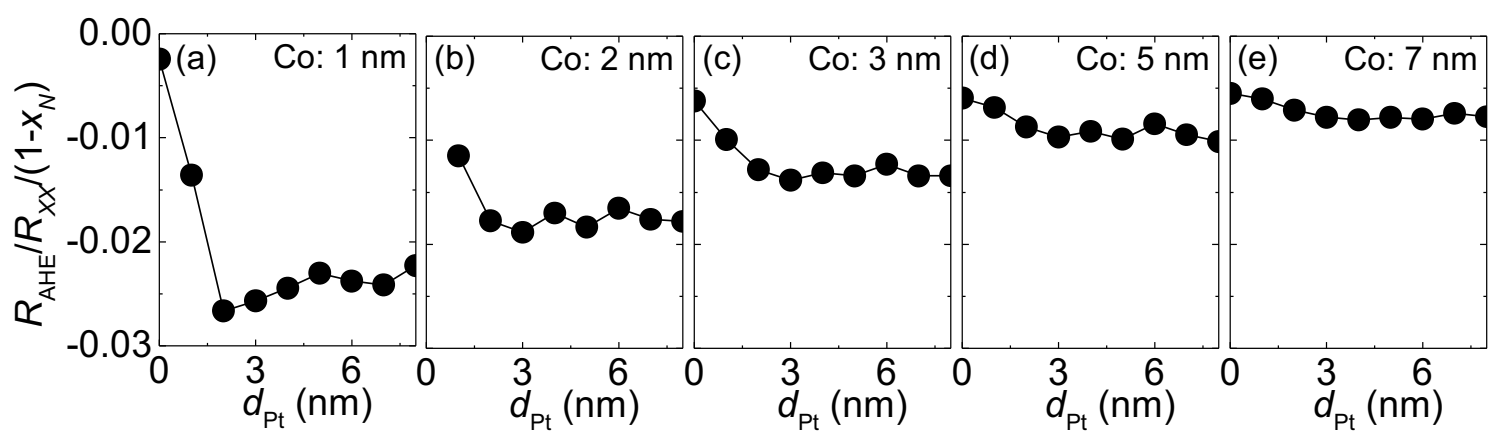

Fig. S3

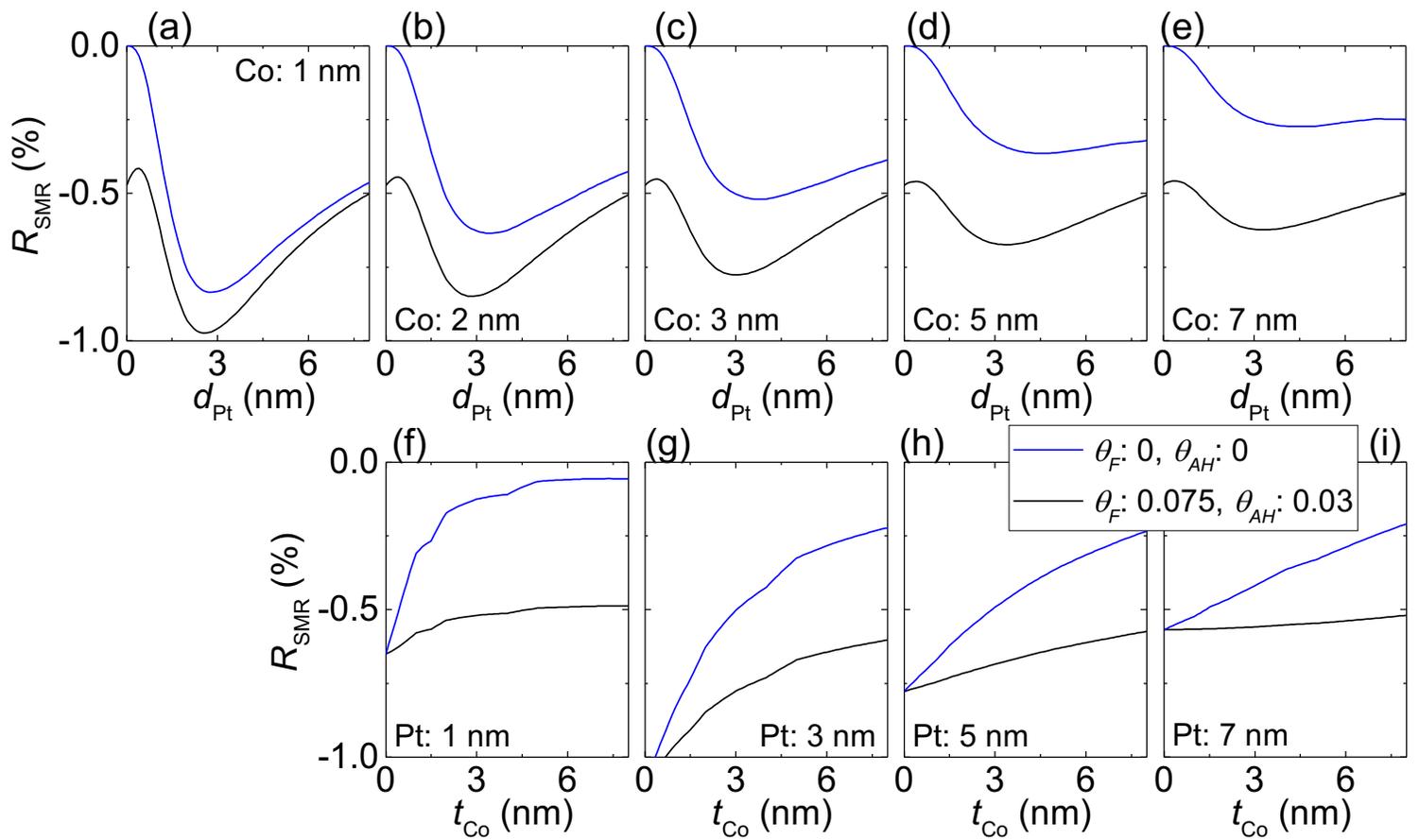

Fig. S4